\documentclass[5p]{elsarticle}
\biboptions{sort&compress}

\usepackage{amssymb}
\usepackage{amsmath}
\usepackage{latexsym}
\usepackage{graphicx}
\usepackage{pifont}
\usepackage{color}

\newcommand{\req}[1]{Eq.\,(\ref{#1})}
\newcommand{\beqn}{\begin{equation}}
\newcommand{\eeqn}{\end{equation}}
\newcommand{\muct}{\tilde\mu_t}

\begin{document}
\title{Higgs two-gluon decay and the top-quark chromomagnetic moment}
\author{Lance Labun and Johann Rafelski}
\address{Department of Physics, University of Arizona, Tucson, Arizona, 85721 USA, and\\
TH Division, Physics Department, CERN, CH-1211 Geneva 23, Switzerland}

\date{9 April, 2012}  

\begin{abstract} 
\vskip-5.8cm
\noindent\hfill CERN-PH-TH/2012-265
\vskip5.3cm 
We study the top quark chromomagnetic factor $\kappa_t$-dependence of the effective interaction  ${\cal L}_{hgg}$ and apply the result to the case of Higgs two gluon decay rate  $\Gamma_{h\to gg}$. Perturbative standard model provides  $\kappa_t< 2$, and for the standard model value $\Gamma_{h\to gg}(\kappa_t)$ is suppressed by 9\% as compared to $\kappa_t= 2$. We further show that in the leading order $\Gamma_{h\to gg}$ for  $\kappa_t\simeq \pm 1.2$ can be very small due to  complete cancelation between the top and bottom quark fluctuations.  We discuss the extension of our results to $\kappa_t>2$ and consider potential paths to obtain experimental information for the $h\to gg$ rate.
\end{abstract}

\begin{keyword}
\PACS 14.65.Ha\sep 14.80.Bn \sep 13.30.Eg \sep 14.70.Dj 
\end{keyword}

\maketitle

{\bf Introduction:}
The Higgs-two-gluon coupling is an effective interaction originating in vacuum fluctuations of the top quark  and to a lesser extent bottom quark. Consequently, the  two gluon decay $h\to gg$  probes physical  properties of the top quark, the heaviest standard model (SM) particle.  The top chromomagnetic moment  holds special interest because of the top's generally-recognized sensitivity to beyond standard model (BSM) physics~\cite{Bernreuther:2008ju} and magnetic moments being a probe of non-pointlike character of quarks~\cite{Brodsky:1980zm}.  Both chromomagnetic and magnetic dipole moments are also sensitive to BSM input from physics at a mass scale beyond $m_t$~\cite{Choudhury:2012np}.

For a pointlike spin-1/2 top quark in the heavy quark limit $2m_t\gg m_h$, the $h\to gg$ amplitude and decay rate are derived from the effective Lagrangian
(see Eq.\,(3.24)~\cite{Dawson:1990zj}, as well as~\cite{Shifman:1979eb} and~\cite{Spira:1995rr})
\beqn\label{lowELeff}
{\cal L}_{hgg}=
\frac{-b_0(\kappa_t)}{1+\gamma_t}\frac{\alpha_s}{2\pi}\frac{h}{v}\frac{1}{4}{\rm Tr}\:G^{\mu\nu}G_{\mu\nu}
\eeqn
where $b_0$ is the QCD $\beta$-function coefficient arising from the top contribution to vacuum polarization, well-known for the case that the top chromomagnetic factor $\kappa_t=2$.  $\gamma_t$ is the top quark anomalous dimension. $h$ is the dynamical Higgs field, $v=246.2~{\rm GeV}$ the Higgs vacuum expectation value, $\alpha_s=g_s^2/4\pi\simeq 0.108 $ the QCD  coupling constant at top-quark scale, and $G_{\mu\nu}\equiv G_{\mu\nu}^at^a$ the gluon field strength tensor. 

In this work, we obtain  $b_0(\kappa_t\ne 2)$ in \req{lowELeff} explicitly as a function of the top quark chromomagnetic ratio and discuss the effect $b_0(\kappa_t\ne 2)$ has on the leading order $h\to gg$ decay amplitude.  The anomalous dimension $\gamma_t$, also a function of $\kappa_t$, is a next-to-leading-order correction and its  $\kappa_t$ dependence will not be considered here. More generally, since \req{lowELeff} does not incorporate all NLO effects~\cite{Spira:1995rr}, we leave comprehensive discussion of NLO corrections to future work.

The chromomagnetic moment $\tilde\mu_t$ is parameterized by the dimensionless chromomagnetic factor $\kappa_t$, 
\beqn\label{mutdefn}
\muct=\frac{\kappa_t}{2}\frac{g_s}{2m_t},
\eeqn 
where $g_s$ is the QCD coupling and $m_t=173.5~{\rm GeV}$ is the top mass.  $\muct$ is defined paralleling the QED magnetic moment, assigning to $\kappa_t$ the role of a chromomagnetic  ``$g$-factor.''  Correspondingly, the top described by the Dirac Lagrangian has implicitly  $\kappa_t=2$ at tree level, as is shown by squaring the Dirac operator $\gamma_5\gamma^{\mu}\Pi_{\mu}=\gamma_5\gamma^{\mu}(i\partial_{\mu}+g_st^aA^a_{\mu})$ with $\Pi$ the Hermitian momentum operator comprising  minimal gauge coupling in the covariant derivative. 

The high top mass implies  unit strength of the top-Higgs minimal coupling, comparable to the QCD coupling $g_s\simeq 1.16$ at this energy scale. The magnitude of these couplings explains the  relatively  large SM perturbative modifications of $\kappa_t$.  In view of the incomplete knowledge about nonperturbative effects and possibility of BSM physics, we consider the chromomagnetic factor  $\kappa_t$  a parameter that can be determined from experiment~\cite{Atwood:1994vm,Haberl:1995ek,Hioki:2009hm,Kamenik:2011dk} and from theoretical studies~\cite{Hewett:1993em,Martinez:1996cy,Martinez:2007qf}.

\begin{figure}[t]
\begin{picture}(230,140)
\put(20,10){\includegraphics[width=0.25\textwidth,angle=90]{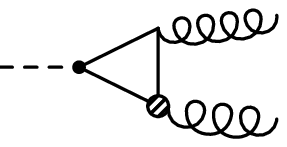}}
\put(47,65){$t$}
\put(40,29){$h$}
\put(43,-2){(a)}
\put(100,10){\includegraphics[width=0.25\textwidth,angle=90]{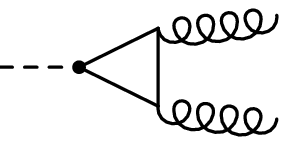}}
\put(127,65){$b$}
\put(123,-2){(b)}
\put(183,10){\includegraphics[width=0.25\textwidth,angle=90]{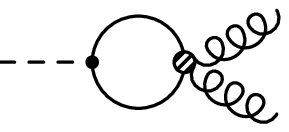}}
\put(207,70){$t$}
\put(203,-2){(c)}
\end{picture}
\caption{The dominant leading-order diagrams generating the effective Higgs-to-two gluon $hgg$ coupling: a)  top and b) bottom quark loops.  One top-gluon nonperturbative vertex has a shaded circle to signify that we consider a general value of the top chromomagnetic moment. c) The 2nd order theory additional diagram including the two top two gluon vertex. \label{fig:diags}}
\end{figure}

\vskip 0.2cm 
{\bf Top quark $\mathbf{b_0(\kappa_t)}$ function:} Figure~\ref{fig:diags} exhibits the diagrams describing top and bottom quark fluctuations leading to the Higgs-two gluon effective coupling.  The top quark contribution is related to its contribution to the renormalization group $\beta$-function of the QCD coupling~\cite{Shifman:1979eb,Spira:1995rr}, manifested in \req{lowELeff} by $b_0$ being leading order coefficient in the loop expansion of the $\beta$-function,
\beqn\label{betadefn}
\beta\equiv\lambda\frac{\partial \alpha}{\partial\lambda}, \quad
\beta(\alpha)=-\frac{b_0}{2\pi}\alpha^2+...
\eeqn 
We take the bottom quark contribution as given by the SM, considering modification of the bottom chromomagnetic moment likely to be smaller.  For this reason, only standard vertices appear in figure \ref{fig:diags}(b).

The functional dependence of $b_0$ on the magnetic moment has been obtained in QED  within the framework of the second-order theory of fermions, both perturbatively~\cite{AngelesMartinez:2011nt} and non-perturbatively~\cite{Rafelski:2012ui}. The advantage of this method is that the (anomalous) magnetic moment interaction is described by the dimension-4 operator and all results we consider here are finite and free of counter terms.  A second-order theory has been discussed for calculating helicity amplitudes including top anomalous dipole moments, and Eq.\,(27) of~\cite{Larkoski:2010am} makes it clear that relating the second order theory to the first order (Dirac) theory requires at least a non-perturbative summation within QCD.

In the second order theory the top is described by the (effective) Lagrangian
\beqn\label{2ndordertheory}
{\cal L}_{\rm top}= \bar\psi\left(\Pi^2-m_t^2-\frac{\kappa_t}{2}\frac{g_s G^{\mu\nu}\sigma_{\mu\nu}}{2}\right)\psi
\eeqn
where $\sigma_{\mu\nu}=(i/2)[\gamma_{\mu},\gamma_{\nu}]$. The Feynman rules of the 2nd order theory~\req{2ndordertheory} include a two-top to two-gluon vertex and hence an additional diagram (c) in figure \ref{fig:diags}.  Upon integrating out the top quark, the dimensionless renormalization parameters $b_0,\gamma_t$ of the reduced theory depend only on the dimensionless top quark point-like couplings, its color charge, and chromomagnetic ratio $\kappa_t$.  We do not consider  here a chromoelectric dipole which can  be  included in \req{2ndordertheory}, since it implies CP violation. 

Considerable simplification in evaluation of $\kappa_t$ dependence is achieved using the low energy theorem \req{lowELeff}~\cite{Shifman:1979eb,Spira:1995rr}.  This is an example of the decoupling theorem for heavy flavor~\cite{Bernreuther:1983}, which implies that below threshold for top pair production $m_h\ll 2m_t$, the $h\to gg$ amplitude can depend in leading order only on the reduced theory arising from integrating out the top. Corrections are in the form of an expansion in $1/m_t$~\cite{Bernreuther:1983} and can be computed separately.  Note that in the 2nd-order theory the chromomagnetic moment is incorporated as a dimension 4 operator and hence is leading order rather than order $1/m_t$, side-stepping naive expectation of decoupling theorems.

When integrating out the top, the 2nd-order formalism \req{2ndordertheory} is necessary because the field has mass dimension 1 and the magnetic moment operator with dimensionless coefficient $\kappa_t/2$ is renormalizable.  This permits us to study large $\kappa_t-2$ and evaluate vacuum fluctuations to any order in the loop expansion.  If we were to attempt a similar computation in the 1st order theory, we encounter the problem that an anomalous chromomagnetic dipole operator is higher dimensioned and nonrenormalizable.  As a result, examples of integrating out the fermion fields in 1st-order formulation of QED with anomalous magnetic moment produced inconsistent, i.e. scheme-dependent results~\cite{Lav85,PauliTerm,Diet78}. 

\begin{figure}
\includegraphics[width=0.47\textwidth]{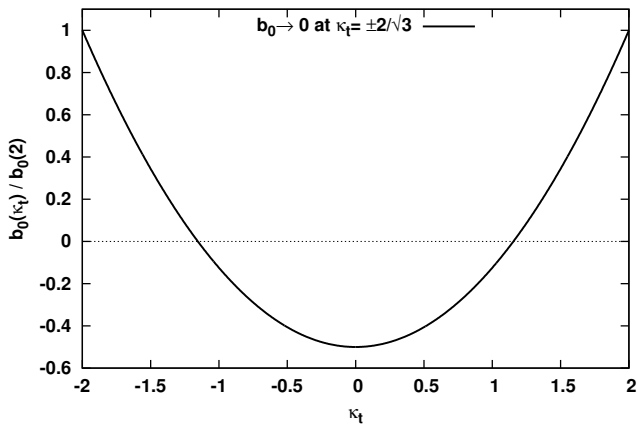}
\caption{The $b_0$ coefficient of the QCD $\beta$ function contributed by the top quark normalized to its value at $\kappa_t=2$. \label{fig:beta}}
\end{figure}

The external field method~\cite{Nielsen:1978rm,Peskin:1995ev} offers a fast track to the generalization of the QED $\beta$-function to QCD, and it provides a result nonperturbative in $\kappa_t$ that agrees with the perturbative result for $|\kappa_t|\le 2$.  At one fermion loop, that calculation is entirely parallel to the one performed in QED.  One need only to modify the QED result by introducing a factor ${\rm tr}(t^at^b)=\delta^{ab}/2$, arising as the trace of two Gell-Mann matrices at the top-gluon vertices. Following the steps in~\cite{Nielsen:1978rm}, one arrives at
\beqn\label{betafunc}
b_0(\kappa_t) =-\frac{2}{3}\left(\frac{3}{8}\kappa_t^2-\frac{1}{2}\right),\  |\kappa_t|\le 2
\eeqn
The $-2/3$ factor separated in front is the well-known value of $b_0$ for $\kappa_t=2$.  \req{betafunc}, normalized by $b_0(2)=-2/3$, is shown in Fig.~\ref{fig:beta}.  

Functional dependence of $b_0$ on  the magnetic moment comparable to \req{betafunc} has been obtained in the QED calculation in perturbation theory~\cite{AngelesMartinez:2011nt} and the non-perturbative in $\kappa$ at one loop-level external field method~\cite{Rafelski:2012ui}.  Thus there is no doubt that in the QCD case where $|\kappa_t|\le 2$ is expected, the contribution of the top quark to the effective $h\to gg$ coupling is obtained using \req{betafunc}. Should it be necessary to consider the case $|\kappa_t|>2$ an  extension of \req{betafunc} following Ref.\cite{Rafelski:2012ui,Labun:2012fg} must be considered.  The method of Ref.\cite{Rafelski:2012ui} assures that the vacuum is stable, and as a result, $b_0(\kappa_t)$ is found to be repeating and  periodic outside the domain $-2<\kappa_t<2$.  

We emphasize that any  extension to  $|\kappa_t|>2$ of the form of \req{betafunc} is a physics motivated choice, based on expectation about the low energy properties of the top-QCD vacuum and to be confirmed by experiment.  Seeing that perturbative evaluation within the SM suggests that $\kappa_t-2<0$ (\req{SMkappaminus2} below), we continue with $|\kappa_t|\le 2$ as the interesting case, \req{betafunc}, and refrain from further discussion of  $b_0(\kappa_t)$ for $|\kappa_t|>2$.

Note in figure~\ref{fig:beta} that $b_0(\kappa_t)$ changes sign at $\kappa_t=\pm 2/\sqrt{3}$ and hence is positive for $|\kappa_t|<2/\sqrt{3}$, due to the decreasing strength of the paramagnetic spin term as $\kappa_t$ diminishes~\cite{Reuter:MG13}.  If in fact $|\kappa_t|<2/\sqrt{3}$ the pattern of interference between the top quark and the lighter quarks changes: for all other quarks $b,c...$ the perturbatively evaluated triangle diagram figure~\ref{fig:diags}(b) yields a positive amplitude, opposite in sign to the top loop.  If the top loop contribution is positive, along with all the other quark loops, the overall Higgs-gluon coupling will have the ``wrong sign'', a possibility recently noted may have a destabilizing impact on the SM~\cite{Reece:2012gi}.

\vskip 0.2cm
{\bf $\mathbf{\kappa_t}$ dependent decay width:}
At leading order, the $h\to gg$ decay width is (see Eq.\,(21) of~\cite{Spira:1995rr})
\beqn\label{ggwidth}
\Gamma(h\to gg)= \Big|\!\sum_{q=t,b,...}\!\! F_q\Big|^2\left(\frac{\alpha_s}{4\pi}\right)^2\frac{m_h^3}{2\pi v^2}
\eeqn
where $F_q$ is a computable function for each quark, in general a function of the ratio $x=4m_q^2/m_h^2$.  Calculating the decay rate from ${\cal L}_{hgg}$ and comparing to \req{ggwidth}, one finds the  factor $F_t$ goes into the $\beta$-function coefficient.  
With the Higgs mass at $m_h\simeq 125.5~{\rm GeV}$, evaluating the amplitude with $F_t\to b_0$ means an error of  a few percent relative to the result from the exact loop amplitude~\cite{Shifman:1979eb,Spira:1995rr}.  Thus, although the ratio $m_h^2/4m_t^2\simeq 0.13$ is not especially small, the heavy quark limit allows a good estimate of the leading-order contribution.  
Higher order QCD corrections to \req{ggwidth} are significant, next to leading order diagrams contributing $\sim 65\%$ enhancement~\cite{Spira:1995rr}.  Recent calculations have been extended to next-to-next-to-leading order including soft gluons to next-to-next-to-leading logarithms~\cite{deFlorian:2009hc}.
Electroweak corrections amount to a couple of percent~\cite{Baglio:2010ae}.

The bottom quark vacuum fluctuation  is included, because it makes a contribution to the amplitude opposite in sign to the top, and consequently interference with top quark amplitude is important.  Neglecting charm and lighter quarks, which make tiny quantitative changes compared to bottom quark effect, the total form factor in \req{ggwidth} is
\beqn
\sum_q F_q\simeq F_t+F_b=b_0(\kappa_t)+F(\frac{4m_b^2}{m_h^2})
\eeqn
The $\beta$-function coefficient $b_0$ is given by \req{betafunc} and the bottom contribution is the $x=4m_b^2/m_h^2$ dependent function
(see Eq.\,(21) of~\cite{Shifman:1979eb} or Eq.\,(3) of~\cite{Spira:1995rr}), 
\beqn\label{formfactor}\displaystyle
F(x)= \begin{cases}
x\left(1-\frac{1-x}{4}\left(\ln\frac{1+\sqrt{1-x}}{1-\sqrt{1-x}}-i\pi\right)^2\right) & x<1\\[0.5cm]
x\big((x-1)\arcsin^2(x^{-1/2})-1\big) & x\geq 1
\end{cases}
\eeqn
For $m_b=4.5~{\rm GeV}$, $F_b$ is opposite in sign and about 17 times smaller than the top quark contribution (at $\kappa_t=2$).

\begin{figure}[t]
\includegraphics[width=0.47\textwidth]{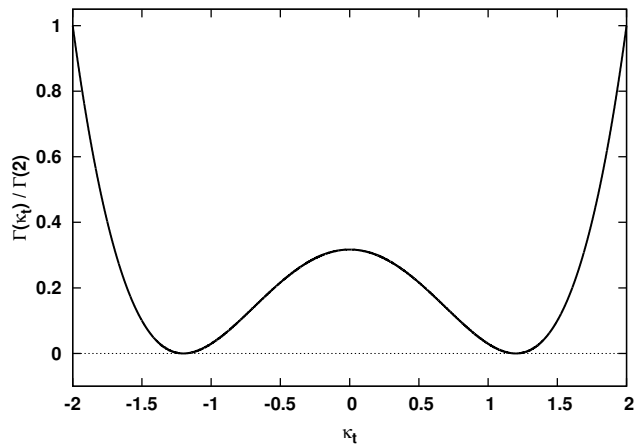}
\caption{Higgs to two gluon $h\to gg$ decay rate normalized to its value at $\kappa_t=2$. }
\label{fig:rate}
\end{figure}

Figure~\ref{fig:rate} shows the $\kappa_t$ dependence of the total  $h\to gg$ decay rate, normalized to its value at $\kappa_t=2$,
\beqn\label{decayrate}
\frac{\Gamma(\kappa_t)}{\Gamma(\kappa_t\to 2)}=\frac{|b_0(\kappa_t)+F(x)|^2}{|b_0(2)+F(x)|^2}
\eeqn
The rate falls quickly for $\kappa_t< 2$ (see below \req{SMkappaminus2}), and for $|\kappa_t|=1.22$ the leading order decay rate goes to zero.  In reality, $\Gamma$ probably does not vanish exactly due to higher order diagrams becoming the dominant contributions---recall that QCD corrections alone are known to be large at next-to-leading order.  Adding the contributions of the other light quarks $c,s...$ would, however, only change the value of $\kappa_t$ at which the leading order $\Gamma \to 0$.

%
\vskip 0.2cm
{\bf Experimental considerations:} 
We believe that the expected magnitude of the decay of Higgs to gluons is significantly modified by $\kappa_t$. Conversely, the measurement of the partial branching ratio of $h\to gg$  would amount to a step towards the measurement of the top-quark chromomagnetic moment. The CMS experiment projects\footnote{See: https://twiki.cern.ch/twiki/bin/view/CMSPublic/ HigProjectionEsg2012TWiki} that the key branching ratios can be measured with a precision that is near 10\% and this encourages further discussion. Computing with $\kappa_t$ as a parameter and obtaining an analytic result allows study of standardized contributions from beyond Standard Model physics and also allows consideration of SM results and constraints derived from other experiments: \\
$\bullet$ The one-loop perturbative SM prediction is (see Eq.\,(4) of~\cite{Martinez:2007qf})
\beqn\label{SMkappaminus2}
\kappa_t-2=-5.6\:10^{-2}
\eeqn
which would lead to a 9\% decrease in the Higgs decay rate.  \\
$\bullet$ Constraints on $\kappa_t$ have been considered already in  the study of  the top quark production~\cite{Atwood:1994vm,Haberl:1995ek,Hioki:2009hm}.  According to Eq.\,(18) of~\cite{Kamenik:2011dk}, present data sets a bound $|\kappa_t-2|<0.2$, which could mean that the decay rate is reduced by up to 25\%.  Considering that many more diagrams (4-gluon, 6-gluon... fusion) contribute to top production, this is just a first step in a more elaborate evaluation of the role of $\kappa_t$ in top production. \\ 
$\bullet$ The radiative decay $b\to s\gamma$ has also been discussed as providing constraints on $\kappa_t$~\cite{Hewett:1993em,Martinez:1996cy}. \\
$\bullet$ Higher order QCD and virtual Higgs can yet make equally important and coherent contributions to the amplitude resulting in a possible major departure of the $h\to gg$ decay rate from prior expectations.  

Having shown how $\kappa_t< 2$ could modify the $h\to gg$  decay rate significantly, we now discuss a means of seeing the effect in experiment.  We argue  that, for Higgs produced at rapidity  $|y|> 2 $, the  $h\to gg$ decay hadron jets  may be separable from randomly correlated directly produced QCD jet  background. Moreover, if $h\to gg$ is visible, it is sure that another reference decay channel with which one can compare is visible as well allowing to measure the relative strength of the glue decay channel.

The two decay gluons produce  two hadron jets back-to-back in the rest frame of the Higgs carrying the significant invariant mass $\simeq 125.5~{\rm GeV}$. While this characteristic may be insufficient to separate the Higgs from the random   hadron jet background at central rapidity,  we suggest to consider events sourced by Higgs particles having rapidity $|y|> 2 $, that is Lorentz factor $\gamma>3.8$. These may be numerous enough given that $\gtrsim 25\%$ of Higgs are expected to be produced with $|y|>2$ according to Fig.\,3 of~\cite{Bozzi:2007pn}. The total energy of the CM frame of back-to-back jets is boosted by $\gamma$  towards and beyond $\gamma m_t\ge 500$ GeV.  Since the transverse momentum of the produced Higgs is expected to be small (a majority of events at $p_T\lesssim 0.3m_H$~\cite{Bozzi:2007pn}), the total transverse momentum of the jets will be on average an order of magnitude  smaller compared to  the boosted longitudinal momentum.  The momentum of $|y|> 2 $ Higgs decay jets is conveniently projected along the collision axis.

Higher order processes contribute to the (incoherent) yield of 4-gluon, 6-gluon, etc. decay events which will depend on the chromomagnetic moment as well.  If the suppression of the 2-gluon decay is large,  further consideration should be given to these multi-jet decays with regard to their sensitivity to the chromomagnetic moment. However, their yield is smaller and the experimental detectability diminishes due to reduced correlation between decay jets creating a higher random jet background.

{\bf Gluon Fusion into Higgs:}
$\kappa_t$ dependence seen in  \req{betafunc} also modifies the rate of Higgs production by gluon fusion.  Moreover, the only effective interaction Higgs coupling with three gluons is within  the form of effective interaction \req{lowELeff}. Thus in the large top mass limit, the top's contribution to the two dominant production processes $gg\to h$ and $gg\to hg$ are both modified by the same pre-factor 
\beqn\label{Production}
\frac{F_t(\kappa_t)}{F_t(2)}=\frac{1+\gamma(2)}{1+\gamma(\kappa_t)}\frac{b_0(\kappa_t)}{b_0(2)}
\eeqn 
Because all production (and decay) processes include exactly one insertion of the effective $hgg$ vertex, they are modified by this same factor at leading order in $m_h/2m_t$.  The inclusive cross-sections for $gg\to h$ and $gg\to hg$ are proportional to the square of \req{Production}.  Measurement of the Higgs+jet(s) cross-sections could thus also offer constraints on $\kappa_t$.

However, we expect that the magnitude of the modification of Higgs production cross section does not follow directly from our present considerations: higher order $2n$-gluon fusion processes with $n>1$ contribute coherently.  The reason $2n$-gluon processes can be important is that the gluon distribution function is steeply rising for small momentum fraction $x$.  It therefore becomes easier to find $n$ gluons with enough total momentum to fuse, since the amount of momentum carried by any single gluon is inversely proportional to $n$. If a large suppression of the higgs-two-gluon decay is found, we would need to revisit the production mechanism providing the observed yield.

The significance of arriving at an accurate measurement of the $h\to gg$ rate, as well as the higgs+jet cross-section  is amplified by our demonstration of their $\kappa_t$-de\-pend\-ence.  In the context of the decoupling theorem~\cite{Bernreuther:1983}, the 2nd-order formulation shows that the leading order $hgg$ coupling can only depend on the top quark charge and chromomagnetic factor.  Our result \req{lowELeff} with $b_0$ given by \req{betafunc} then supplies the definitive characterization of the leading order contribution.

\vskip 0.2cm
{\bf Conclusions:}
The SM prediction is that a scalar Higgs at mass $m_h\simeq 125.5$ GeV decays in about 8.5\% of cases into two parallel polarized gluons~\cite{Denner:2011mq} which turn into hadron jets.  The two gluon decay is dominated by top-quark vacuum loop and thus is sensitive to modifications introduced by the top quark chromomagnetic moment. The top quark vacuum loop also enters the  Higgs two photon decay which situation differs in two key properties: a) the two photon decay branching ratio is a tiny but visible 0.2\% contribution to the total Higgs decay, and, b) the top quark loop plays a subordinate role to the $W$ loop. Moreover, the standard model modification of top quark chromomagnetic moment that enters the two-gluon decay channel differs in quantitative manner from the modification expected for top quark (electro)magnetic moment. This shows that the sensitivity of the two-gluon and two-photon decay processes to top-quark loops, and therefore the experimental opportunity differ.

Although Higgs-two-gluon decay is challenging to measure, our study highlights the fundamental interest in arriving at a precise result for this decay channel, considering the connection made here between the Higgs-two-gluon decay rate and the top quark chromomagnetic moment $\mu_t$. Among today-measurable parameters $\mu_t$ is arguably the most sensitive probe to BSM effects.   For example, a common change in magnitude of the magnetic moment of all quarks \req{mutdefn} would most influence the Bohr magneton of the heaviest quark, since its natural scale $1/m_t$  is smallest by a large factor.   

Higgs production by gluon fusion also involves heavy quark vacuum fluctuations and is subject to modification induced by  $\kappa_t\ne 2$.  In the infinite top mass limit, higgs and higgs-plus-jet production are obtained from \req{lowELeff} and our result modifies the effective coupling equally in the production amplitudes.  When exploring the production process, we discovered that the magnitude of the modification of gluon-Higgs production cross section does not follow directly from our present considerations. A considerably more detailed investigation is needed to understand the influence of  $\kappa_t$ in the realm of Higgs production.

We remark that the reduction in the higgs production and the decay rates resulting from $\kappa_t-2$ could be compensated by fourth generation quarks contributing to the total loop amplitude.  Constraints on the fourth generation from $hgg$ coupling are thus loosened until $\kappa_t$ and the $hgg$ effective coupling are measured independently.

To summarize, the chromomagnetic moment of the top quark offers a well-known opportunity to search for quark compositeness and BSM physics. We have considered the top chromomagnetic moment $\kappa_t$ a parameter, and have shown the dependence of the leading order Higgs-two-gluon interaction ${\cal L}_{hgg}$ on  $\kappa_t$.  Due to the dominant role of the top in the effective Higgs-glue interaction,  we obtain a large sensitivity of the Higgs-two-gluon $h\to gg$ decay to the top chromomagnetic moment: even the one-loop SM estimate of $\kappa_t-2$ leads to a 10\% suppression of the 2-gluon decay channel, and a much larger effect could result from  quark compositeness or other BSM physics.  Conversely, absence of visible effect may push the associated  BSM energy scale up significantly due to the hyper-sensitivity of particle magnetic moment to new physics~\cite{Brodsky:1980zm,Choudhury:2012np}.

\vskip 0.3cm
{\it Acknowledgments:} L.L. and J.R.  thank the  TH-division of the CERN Physics Department for hospitality where this work was initiated.  We thank K. Howe for discussion.  This work was supported by a grant from the  US Department of Energy, DE-FG02-04ER41318.

\end{document}